\documentclass[prx,aps,twocolumn,preprintnumbers,amsmath,amssymb,superscriptaddress,showpacs]{revtex4-1}

\usepackage{amsmath}
\usepackage{graphicx}
\usepackage{amssymb}
\usepackage{color}
\usepackage[dvipdfm,bookmarks=true,colorlinks=true,linkcolor=blue,urlcolor=blue,citecolor=blue]{hyperref}

\begin{document}

\title{Self-consistent spin-wave analysis of the 1/3 magnetization
plateau in the kagome antiferromagnet}

\author{Zhong-Chao~Wei}
\affiliation
{Institute of Physics, Chinese Academy of Sciences, P.O. Box 603, Beijing
100190, China}

\author{Hai-Jun~Liao$^{*}$}
\affiliation
{Institute of Physics, Chinese Academy of Sciences, P.O. Box 603, Beijing
100190, China}

\author{Jing~Chen}
\affiliation
{Institute of Physics, Chinese Academy of Sciences, P.O. Box 603, Beijing
100190, China}

\author{Hai-Dong~Xie}
\affiliation
{Institute of Physics, Chinese Academy of Sciences, P.O. Box 603, Beijing
100190, China}

\author{Zhi-Yuan~Liu}
\affiliation
{Institute of Physics, Chinese Academy of Sciences, P.O. Box 603, Beijing
100190, China}

\author{Zhi-Yuan~Xie}
\affiliation
{Department of Physics, Renmin University of China, Beijing 100872, China}

\author{Wei~Li}
\affiliation
{Department of Physics, Key Laboratory of Micro-Nano Measurement-Manipulation
and Physics (Ministry of Education), Beihang University, Beijing 100191, China}
\affiliation
{International Research Institute of Multidisciplinary Science, Beihang
University, Beijing 100191, China}

\author{Bruce~Normand}
\affiliation
{Department of Physics, Renmin University of China, Beijing 100872, China}

\author{Tao~Xiang}
\affiliation
{Institute of Physics, Chinese Academy of Sciences, P.O. Box 603, Beijing
100190, China}
\affiliation{Collaborative Innovation Center of Quantum Matter, Beijing
100190, China}

\date{\today}

\begin{abstract}
We propose a modified spin-wave theory to study the $1/3$ magnetization
plateau of the antiferromagnetic Heisenberg model on the kagome lattice. By
the self-consistent inclusion of quantum corrections, the $1/3$ plateau is
stabilized over a broad range of magnetic fields for all spin quantum
numbers, $S$. The values of the critical magnetic fields and the widths of
the magnetization plateaus are fully consistent with recent numerical
results from exact diagonalization and infinite projected entangled paired
states.
\end{abstract}


\maketitle

The investigation of low-dimensional frustrated magnetism has become one of
the most active frontiers in condensed matter physics. Current frontiers in
the field include obtaining full insight into the entanglement structure of
quantum many-body wavefunctions for different types of quantum spin liquid
\cite{SpinLiquid} and simplex-solid state \cite{Arovas08}. Among the many
different frustrated systems, quantum antiferromagnets on the kagome lattice
are perhaps the most intriguing, because the combination of strong geometric
frustration and weak constraints maximizes quantum fluctuation effects. The
kagome antiferromagnet has attracted ever-increasing attention over the
last two decades, with many different methods applied and resulting proposals
for the nature of the ground state \cite{VBC_Kagome,Z2SL_Kagome,XCY+14,
U1SL_Kagome}. Many materials realizations of the kagome geoemtry have now
been discovered, including volborthite ($\mathrm{Cu_3V_2O_7(OH)_2 \cdot
2H_2O}$) \cite{Volborthite}, herbertsmithite ($\mathrm{ZnCu_3(OH)_6Cl_2}$)
\cite{Herbertsmithite}, vesignieite ($\mathrm{BaCu_3(OH)_6Cl_2}$)
\cite{vesignieite}, $\mathrm{BaNi_3(OH)_2(VO_4)_2}$ \cite{nickelate},
$\mathrm{KV_3Ge_2O_9}$ \cite{vanadate}, and jarosites of several different
metal ions including chromium ($\mathrm{KCr_3(OH)_6(SO_4)_2}$) \cite{jarosite}.

One of the characteristic features of kagome antiferromagnets is the
appearance of magnetization plateaus in the presence of an external
magnetic field. Irrespective of the method applied and the prediction for
the zero-field ground state, all theoretical approaches agree that there
exists a robust magnetization plateau at $m = 1/3$ for all values of the
spin quantum number, $S$. The 1/3 plateau has been investigated theoretically
by real-space perturbation theory (RSPT) \cite{RSPT}, exact diagonalization
(ED) \cite{ED_data,Richter}, density-matrix renormalization-group methods
(DMRG) \cite{DMRG_data}, and infinite projected entangled paired states
(iPEPS) \cite{iPEPS}. RSPT provides analytical results for the critical
magnetic fields and the width of the plateau \cite{RSPT}. However, a
qualitative discrepancy has arisen with recent numerical results from ED
\cite{ED_data} and iPEPS \cite{iPEPS}, not least in that the calculated
plateau width increases with increasing $S$, whereas it decreases within RSPT.

\begin{figure}[t]
\centering
\includegraphics[width=7.0cm]{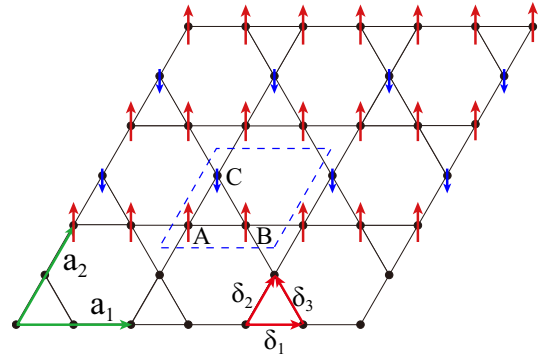}
\caption{The up-up-down spin configuration of the $1/3$ magnetization plateau
on the kagome lattice. The two primitive lattice vectors are denoted as
$\vec{a}_1 = (1,0) \, a$ and $\vec{a}_2 = (\frac{1}{2},\frac{\sqrt 3}{2}) \,
a$, where we set $a = 1$. The unit cell, denoted by the dashed blue lines,
contains the three sites $(A, B, C)$. $\delta_{1}$, $\delta_{2}$, and
$\delta_{3}$ denote the nearest-neighbor lattice vectors.}
\label{Fig:Lattice}
\end{figure}

To improve both qualitative and quantitative understanding of the $1/3$
magnetization plateau in the kagome antiferromagnet, in this Letter we
employ a self-consistent spin-wave theory to study its properties. The
theory contains a single quantum correction parameter, determined
self-consistently from the expectation values of the magnon densities.
We compute these densities for all $S$, derive the spin-wave spectrum,
and evaluate both the critical magnetic fields and the width of the $1/3$
plateau, finding complete consistency with the recent numerical results.

We study the Hamiltonian
\begin{equation}
H = \sum_{\langle i,j \rangle } {\vec S}_{i} \cdot {\vec S}_{j} - h \sum_{i} S_{i}^{z},
\label{eq:Model}
\end{equation}
where ${\vec S}_i$ is the spin-$S$ operator on site $i$, $\langle i,j \rangle$
denotes the sum over neighboring sites, and $J$, the nearest-neighbor
antiferromagnetic exchange coupling, is set as the energy scale ($J = 1$).
Recent numerical studies of the kagome and Husimi lattices by the method
of projected entangled simplex states (PESS) \cite{XCY+14} have demonstrated
very explicitly \cite{Liao+16} that the origin of the $1/3$-plateau phase is
the creation of a semiclassical up-up-down spin configuration on every
triangle, as shown in Fig.~\ref{Fig:Lattice}; we stress that this statement
holds for all values of $S$, even $S = 1/2$. As a consequence, it is entirely
justified to employ a spin-wave description, which we implement by performing
a Holstein-Primakoff transformation from spin operators to bosonic degrees
of freedom,
\begin{eqnarray}
S_{i}^{+} & = & \sqrt{2S - d_{i}^{\dagger} d_{i}} d_{i}, \,\:
S_{i}^{z} = S - d_{i}^{\dagger} d_{i}, \: i \in A,B, \qquad \\
S_{i}^{+} & = & - d_{i}^{\dagger} \sqrt{2S - d_{i}^{\dagger} d_{i}}, \:\:
S_{i}^{z} = d_{i}^{\dagger} d_{i} - S, \: i \in C.
\end{eqnarray}
Here we have assumed that spins on the $A$ and $B$ sublattices are oriented
along ${\hat z}$, whereas those on the $C$ sublattice are oriented along
$-{\hat z}$. We restrict all of our considerations to zero temperature.

In a conventional linear spin-wave theory ($S \rightarrow \infty$), the
$1/3$ plateau is stable only at $h = 2S$. However, the effects of quantum
fluctuations at finite $S$ may be included through the effective mean-field
boson densities at each site, $n_{i} \equiv \langle d_{i}^{\dagger} d_{i}\rangle$,
which are determined self-consistently and act to stabilize the magnetization
plateau over a finite range of $h$. By introducing the approximation
\begin{equation}
\sqrt{2S - d_{i}^{\dagger} d_{i}} \approx \sqrt{2S - \langle d_{i}^{\dagger} d_{i}
\rangle} = \sqrt{2S - n_{i}},
\end{equation}
the Hamiltonian (\ref{eq:Model}) can be decoupled in the form
\begin{eqnarray}
\mathcal{H}_{\rm MF} = \sum_{k} \alpha_{k}^{\dagger} \left[\mathcal{H}
       (\boldsymbol{k}) + \delta h \Lambda \right] \alpha_{k} + \mathcal{C},
\label{eq:HMF}
\end{eqnarray}
where $\alpha_{k}\equiv(a_{k},b_{k},c_{-k}^{\dagger})^{T}$, $a_{k}$, $b_{k}$, and
$c_{k}$ are the Fourier transforms of the operators, $d_{i}$, for each of the
$A$, $B$, and $C$ sublattices, $\delta h = (h - 2S)$, $\Lambda =
\mathrm{diag}(1,1,-1)$, and $\mathcal{C}$ is a constant. $\mathcal{H}
(\boldsymbol{k})$ specifies the quadratic Hamiltonian at $h = 2S$,
\begin{equation*}
\mathcal{\hat{H}}(\boldsymbol{k}) = \left(\begin{array}{ccc}
   2S  & d_{k} & f_{k} \\ d_{k} &   2S   & g_{k} \\ f_{k} & g_{k} &   2S
\end{array}\right) \!,
\end{equation*}
in which
\begin{eqnarray}
d_{k} & = & \quad \sqrt{(2S - n_{A})(2S - n_{B})} \, \cos(\boldsymbol{k} \cdot
\boldsymbol{\delta_{1}}), \nonumber \\
f_{k} & = & - \sqrt{(2S - n_{A})(2S - n_{C})} \, \cos(\boldsymbol{k} \cdot
\boldsymbol{\delta_{2}}),\\
g_{k} & = & - \sqrt{(2S - n_{B})(2S - n_{C})} \, \cos(\boldsymbol{k} \cdot
\boldsymbol{\delta_{3}}), \nonumber
\label{edfg}
\end{eqnarray}
with the bond vectors $\delta_{i}$ as shown in Fig.~\ref{Fig:Lattice}.

In Eq.~(\ref{edfg}), $n_{A}$, $n_{B}$, and $n_{C}$ are respectively the
expectation values of the magnon density, $\langle d_{i}^{\dagger} d_{i}\rangle$,
on each of the $A$, $B$, and $C$ sublattices. From the Holstein-Primakoff
transformation, the normalized longitudinal magnetizations on the three
sublattices are $m_A = 1 - n_A/S$, $m_B = 1 - n_B/S$, and $m_C = n_C/S - 1$.
In the $1/3$-plateau phase, by definition $(m_A + m_B + m_C)/3 = 1/3$ and
therefore $n_A + n_B = n_C$. Further, here and henceforth we employ the
reflection symmetry of the system about a vertical axis through the C sites
[Fig.~\ref{Fig:Lattice}], which specifies that $m_A = m_B$ and $n_A = n_B$.
Thus the self-consistent spin-wave theory for the 1/3-plateau phase contains
only one parameter to be determined, which we denote as $x = n_{A} = n_{B} =
n_{C}/2$.

The diagonalization of a general quadratic bosonic Hamiltonian is nontrival.
Here we summarize the procedure \cite{MLM13} to diagonalize the mean-field
Hamiltonian of Eq.~(\ref{eq:HMF}). In the generalized Bogoliubov
transformation
\begin{equation}
\alpha_{k} = U_{k} \beta_{k},
\label{Bogoliubov}
\end{equation}
the commutation relations of the new bosons, $\beta_{k} \equiv (\beta_{1,k},
\beta_{2,k}, \beta_{3,-k}^{\dagger})^{T}$, are preserved if $U_{k}$ satisfies the
condition
\begin{equation}
U_{k}^{\dagger} \Lambda U_{k} = U_{k} \Lambda U_{k}^{\dagger} = \Lambda.
\label{Unitary_Relations}
\end{equation}
By substituting Eq.~(\ref{Bogoliubov}) back into Eq.~(\ref{eq:HMF}) and
making use of the condition (\ref{Unitary_Relations}), one obtains
\begin{equation}
\mathcal{H}_{MF} = \sum_{k} \beta_{k}^{\dagger} \left[ U_{k}^{\dagger}
\mathcal{H} (\boldsymbol{k}) U_{k} + \delta h \Lambda \right] \beta_{k}
 + \mathcal{C}.
\end{equation}

To obtain a diagonal form, $U_{k}$ must satisfy the further condition
\begin{equation}
U_{k}^{\dagger} \mathcal{H} (\boldsymbol{k}) U_{k} = D_{k} = \mathrm{diag}
(\lambda_{1} (\boldsymbol{k}), \lambda_{2} (\boldsymbol{k}),
\lambda_{3} (\boldsymbol{k})),
\label{DiagonalCondition}
\end{equation}
where the eigenvalues $\{\lambda_{n}(\boldsymbol{k})\}$, which are the
spin-wave dispersion relations at $h = 2S$, should be positive definite,
i.e.~$\lambda_{n} (\boldsymbol{k}) > 0$. This requires that the matrix
$\mathcal{H}(\boldsymbol{k})$ also be positive definite \cite{MLM13}, in
which case there exists a matrix, $K$, such that $\mathcal{H} (\boldsymbol{k})
= K^{\dagger} K$ (Cholesky decomposition or eigendecomposition of $\mathcal{H}
(\boldsymbol{k})$). The diagonalization of $\mathcal{H} (\boldsymbol{k})$
therefore maps to the diagonalization of $K \Lambda K^{\dagger}$, meaning to
the exercise of finding a further unitary matrix, $V$, such that $V^{\dagger}
(K \Lambda K^{\dagger}) V = L$, with $L$ diagonal. The solutions satisfying
the two conditions (\ref{Unitary_Relations}) and (\ref{DiagonalCondition})
simultaneously are then $D_k = \Lambda L$ and $U_{k} = K^{-1} V D^{1/2}_{k}$.
$U_{k}$ may in fact be obtained directly by diagonalizing the matrix
$\Lambda \mathcal{H}(\boldsymbol{k})$, i.e.~from the equation
$(\Lambda \mathcal{H}(\boldsymbol{k})) U_{k} = U_{k} (\Lambda D_{k})$.

It is important to note that, due to Eq.~(\ref{Unitary_Relations}), the
generalized Bogoliubov transformation matrix, $U_{k}$, is independent of
the magnetic field, $h$. Thus the energies of the spin-wave excitations at
fields away from $h = 2S$ are obtained simply by uniform shifts of the
three magnon modes at $h = 2S$,
\begin{equation}
\omega_{1,2} (\boldsymbol{k}) = \lambda_{1,2} (\boldsymbol{k}) + \delta h,
\quad \omega_{3} (\boldsymbol{k}) = \lambda_{3} (\boldsymbol{k}) - \delta h,
\label{Spectrum}
\end{equation}
whence the mean-field Hamiltonian (\ref{eq:HMF}) can be rewritten
\begin{equation}
\mathcal{H}_{MF} = \sum_{k} \sum_{j=1}^{3} \omega_{j} (\boldsymbol{k})
\beta_{j,k}^{\dagger} \beta_{j,k} + \mathcal{C}.
\end{equation}

In order to show the spin-wave dispersion relations, $\omega_{j}
(\boldsymbol{k})$, it is necessary to solve for $x$. However, some
preliminary remarks on the nature of the 1/3-plateau phase are already
in order. It is clear from Eq.~(\ref{Spectrum}) that modes $\omega_{1,2}
(\boldsymbol{k})$ are pushed up by an increase of the magnetic field while
mode $\omega_{3} (\boldsymbol{k})$ is pushed down, and conversely when $h$
decreases. When the lowest mode touches energy zero, the plateau phase
becomes unstable. Thus the lower transition point, $h_{c1}$, out of the
$1/3$-plateau phase is determined by the lower gap of two modes $\lambda_{1,2}
(\boldsymbol{k})$, whereas the upper transition, $h_{c2}$, is determined by
the gap of $\lambda_{3}(\boldsymbol{k})$. Defining $\Delta_{j} = \mathrm{min}
(\lambda_{j}(\boldsymbol{k}))$ as the energy gaps of the three spin-wave
branches at field $h = 2S$,
\begin{equation}
h_{c1} = 2S - \mathrm{min} (\Delta_{1},\Delta_{2}), \quad h_{c2} = 2S + \Delta_{3},
\label{eq:CriticalFields}
\end{equation}
and the width of the plateau is given by
\begin{eqnarray}
\Delta_{w} = h_{c2} - h_{c1} = \mathrm{min} (\Delta_{1},\Delta_{2}) + \Delta_{3}.
\label{Plateau_Width}
\end{eqnarray}

To determine the magnon-density parameter, $x$, we use a numerical iterative
method to solve the equation
\begin{eqnarray}
x = \frac{1}{N} \sum_{k} \left|U_{k}(1,3)\right|^{2},
\label{SCEQ}
\end{eqnarray}
where $U_{k}$ is the Bogoliubov transformation matrix of Eq.~(\ref{Bogoliubov}),
which depends in turn on $x$. Because $U_{k}$ does not depend on the magnetic
field, $x$ is also independent of $h$ in the 1/3-plateau phase, although it
does depend on $S$. As noted above, we need therefore solve Eq.~(\ref{SCEQ})
only at $h = 2S$ to obtain the spin-wave spectra, $\{ \omega_{j}
(\boldsymbol{k}) \}$, and hence the critical fields $h_{c1}$ and $h_{c2}$
(\ref{eq:CriticalFields}).

\begin{figure}[t]
\centering
\includegraphics[width=8.0cm]{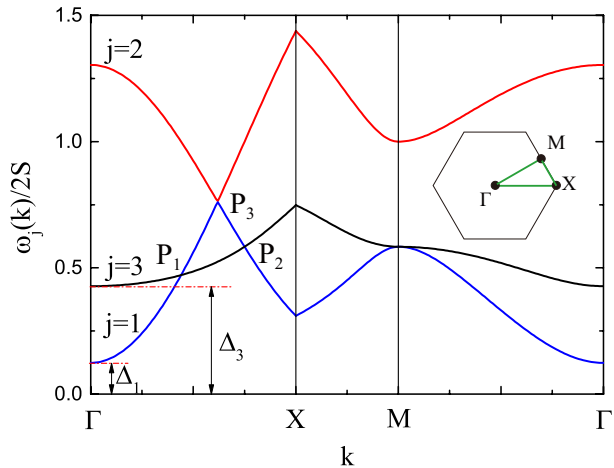}
\caption{Dispersion relations of the three elementary magnon excitations,
$j = 1$, 2, and 3, for selected symmetry directions in the first Brillouin
zone (inset), illustrated for field $h = 2S$ and $S = 1/2$. $\Delta_{1}$ and
$\Delta_{3}$ denote respectively the energy gaps of modes $1$ and $3$.}
\label{Fig:Dispersions}
\end{figure}

The spin-wave spectra for the $S = 1/2$ case at field $h = 2S$ are shown in
Fig.~\ref{Fig:Dispersions}. The minima of the two lower modes, $j = 1$ and 3,
are both located at the $\Gamma$ point ($\boldsymbol{k = 0}$). There are two
trivial level-crossings between these modes, occurring at the points marked
$P_1$ and $P_2$ in Fig.~\ref{Fig:Dispersions} but actually forming a circle
in the Brillouin zone, which changes size with $\delta h$. However, the
crossing between modes $j = 1$ and 2, which have the same $\delta
h$-dependence, occurs at a single point, $P_3$, along the line $\Gamma$X.
This nontrivial exact crossing is a consequence of the reflection symmetry
through the ${\hat y}$-axis and leads to a Dirac-type spectrum between the
eigenmodes $j = 1$ and 2. In more detail, the magnon spectra along $\Gamma$X
can be found analytically by diagonalizing the matrix $\Lambda \mathcal{H}
(\boldsymbol{k}=(k_x,0))$ to obtain
\begin{eqnarray}
\!\!\!\! \lambda_{1}(k_x) & = & \begin{cases}
2S - (2S \! - \! x) \cos (k_x/2), \quad \;\; & \!\!\! k_x \le k_c \\
\frac{1}{2} [\sqrt{\gamma_{k}} + (2S \! - \! x) \cos (k_x/2)], & \!\!\!
k_x > k_c \\ \end{cases} \nonumber \\
\!\!\!\! \lambda_{2}(k_x) & = & \begin{cases} \frac{1}{2} [\sqrt{\gamma_{k}} +
(2S \! - \! x) \cos (k_x/2)], & \!\!\! k_x \le k_c \\ 2S - (2S \! - \! x)
\cos (k_x/2), & \!\!\! k_x > k_c \\
\end{cases} \\
\!\!\!\! \lambda_{3}(k_x) & = & \;\;\: {\textstyle \frac{1}{2}}
[\sqrt{\gamma_{k}} - (2S \! - \! x) \cos (k_x/2)], \nonumber
\end{eqnarray}
with $\gamma_{k} = [(2S - x) \cos (k_x/2) + 4x]^{2} + 24 x (S - x)$. The
crossing point is therefore located at momentum $k_x = k_c = 2 \cos^{-1} \left(
\frac{3 + x/S - \sqrt{1 + 18(x/S) - 3(x/S)^2}}{4 - 2(x/S)} \right)$. In the
classical limit, $S \rightarrow \infty$, the position of crossing is $k_c =
2\pi/3$. The gaps of modes $1$ and $3$ are
\begin{equation}
\Delta_{1} = x, \quad \Delta_{3} = \frac{x}{2} - S + \sqrt{S^{2} + 9 S x -
\frac{15}{4}x^{2}},
\label{Gap13}
\end{equation}
and hence the normalized critical magnetic fields and the width of the
$1/3$ plateau are
\begin{eqnarray}
\frac{h_{c1}}{S} & = & 2 - \frac{x}{S}, \label{ehc1} \\
\frac{h_{c2}}{S} & = & 2 - \! \left( \! 1 - \frac{x}{2S} \right) \! + \sqrt{1
+ 9 \! \left( \frac{x}{S} \right) \! - \frac{15}{4} \! \left( \frac{x}{S}
\right)^{2}}, \label{ehc2} \\
\frac{\Delta_{w}}{S} & = & \frac{3}{2} \left( \frac{x}{S} \right) - 1 +
\sqrt{1 + 9 \left( \frac{x}{S} \right) - \frac{15}{4} \! \left( \frac{x}{S}
\right)^{2}}.
\label{edw}
\end{eqnarray}

\begin{figure}[t]
\centering
\includegraphics[width=8.0cm]{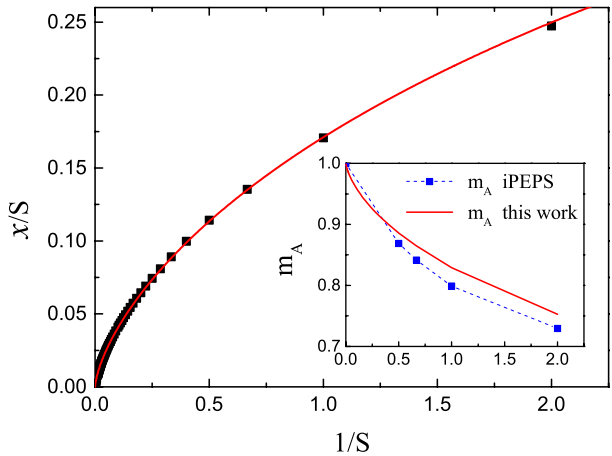}
\caption{Self-consistent magnon-density parameter, $x/S$, shown as a function
of $1/S$. The red line indicates a two-parameter power-law fit of the form
$x/S = \mu (1/S)^{\eta} / [1 + \mu(1/S)^{\eta}]$ [Eq.~(\protect{\ref{ex}})].
The inset shows the magnitudes of the normalized magnetization on the $A$ and
$B$ sublattices, with the result $m_A = m_B = 1 - x/S$ from the self-consistent
spin-wave analysis shown as the solid red line and the corresponding results
from iPEPS calculations with bond dimension $D = 10$ \cite{iPEPS} shown as
square blue points.}
\label{Fig:Parameter}
\end{figure}

Concerning the form of the magnon-density function, $x(S)$, in
Fig.~\ref{Fig:Parameter} we show the quantity $x/S$ determined numerically
by solving the self-consistent equation (\ref{SCEQ}). It is clear that
$x/S$ grows sub-linearly from zero, and similarly that $m_A = m_B = 1 - x/S$
(inset, Fig.~\ref{Fig:Parameter}) falls monotonically from full polarization
in the classical limit. We propose the power-law form
\begin{equation}
m_A = m_B = \frac{1}{1 + \mu (1/S)^\eta}
\label{emab}
\end{equation}
for the sublattice magnetization and thus
\begin{equation}
\frac{x}{S} = \frac{\mu(1/S)^{\eta}}{1 + \mu (1/S)^{\eta}} \, = \, \frac{1}{1 +
\mu^{-1} S^\eta}
\label{ex}
\end{equation}
for the magnon density. We find that this two-parameter fit, with prefactor
$\mu = 0.206(1)$ and exponent $\eta = 0.690(1)$, offers an excellent account
of the data over the entire range of $S$, i.e.~not only where $1/S$ is small
but even when $S = 1/2$ (last point). We comment that significantly better
statistics still can be obtained by generalizing this type of power-law fit
to two exponents.

\begin{figure}[t]
\centering
\includegraphics[width=8.0cm]{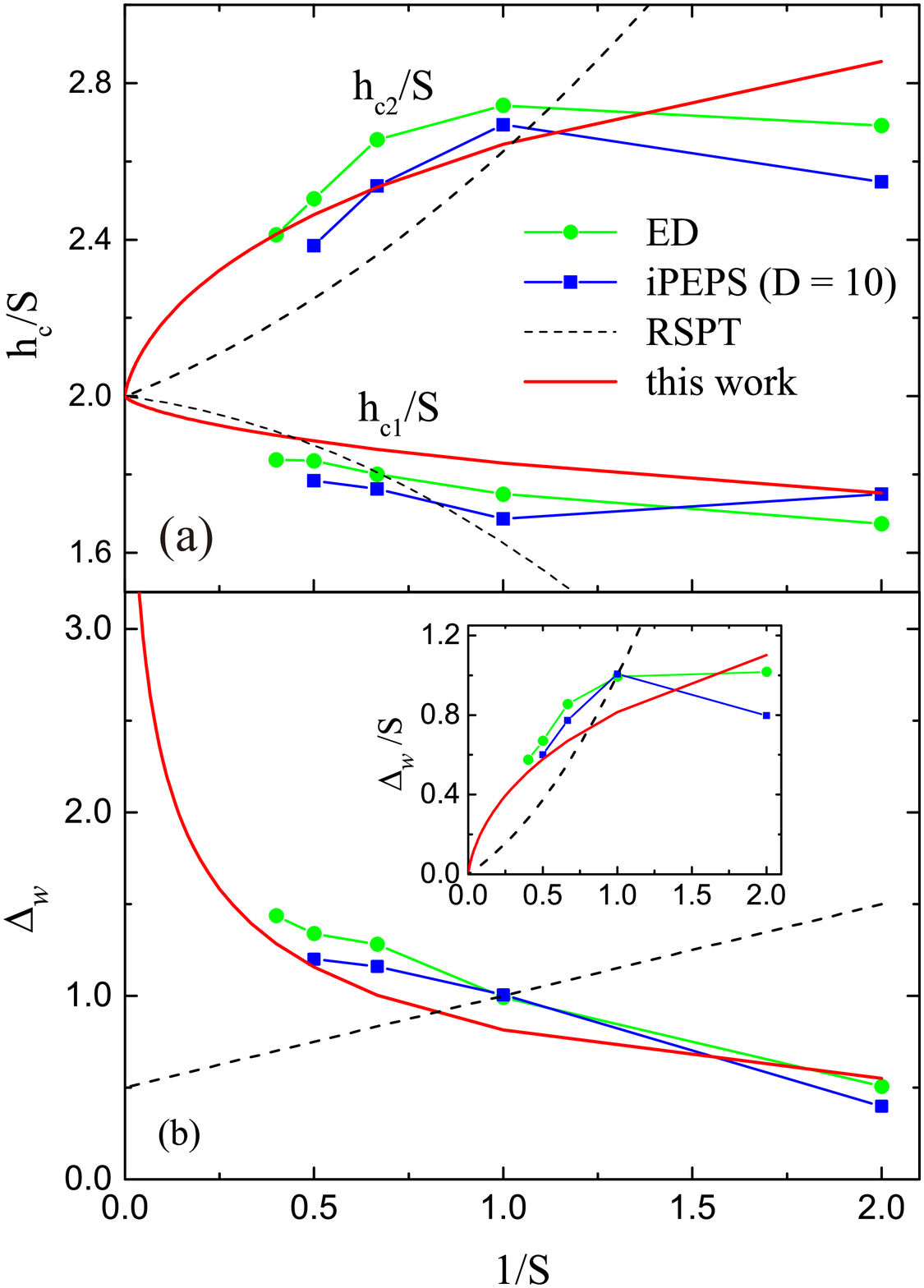}
\caption{(a) Normalized critical magnetic fields, $h_{c1}/S$ and $h_{c2}/S$,
of the $1/3$-plateau phase, shown as functions of $1/S$. Data from the
self-consistent spin-wave theory (solid red lines) are compared with results
from ED (green) \cite{ED_data}, iPEPS with bond dimension $D = 10$ (blue)
\cite{iPEPS}, and RSPT (dashed black) \cite{RSPT}; (b) Width $\Delta_{w} =
(h_{c2} - h_{c1})$ of the $1/3$ plateau, shown as a function of $1/S$; all
data labels as in panel (a). The inset shows the normalized width,
$\Delta_{w}/S$.}
\label{Fig:CriticalFields}
\end{figure}

Returning to Eqs.~(\ref{ehc1}) to (\ref{edw}), the nonzero boson-density
expectation values on all sites ($n_A = n_B = n_C/2 = x$) lead to finite
energy gaps [Eq.~(\ref{Gap13})], which stabilize the 1/3 plateau over a
broad range of magnetic fields, as shown in Fig.~\ref{Fig:CriticalFields}
for all values of $S$. In the self-consistent spin-wave treatment, all of
these quantities increase with the spin quantum number, $S$, although
their normalized values decrease towards the expected limits
(Figs.~\ref{Fig:Parameter} and \ref{Fig:CriticalFields}). This type of
behavior is completely consistent with the numerical results obtained from
ED \cite{ED_data} and iPEPS \cite{iPEPS}, which are shown for comparison in
Fig.~\ref{Fig:CriticalFields}. The $1/S$-dependence of both critical fields,
$h_{c1}$ and $h_{c2}$, and of the plateau width, $\Delta_{w}$, predicted by the
self-consistent spin-wave theory have the correct qualitative trends, and in
fact close quantitative agreement with the numerical calculations. It is
striking that our analytical results are accurate at a semi-quantitative
level even for the extreme quantum case, $S = 1/2$. These results demonstrate
that the self-consistent spin-wave theory captures properly the nature of
the $1/3$ magnetization plateau in the kagome antiferromagnet.

By contrast, in the RSPT approach \cite{RSPT} one performs an expansion in
powers of $1/S$ to obtain the plateau properties in the form
\begin{eqnarray}
h_{c1}/S & = & 2 - \frac{1}{8S} - \frac{1}{4S^{2}}, \\
h_{c2}/S & = & 2 + \frac{3}{8S} + \frac{1}{4S^{2}}, \\
\Delta_{w}/S & = & \frac{1}{2S} + \frac{1}{2S^{2}},
\end{eqnarray}
to order $(1/S)^2$. However, not only is this form fated to diverge in the
most quantum systems, as $1/S \rightarrow 1$, but there is also a qualitative
discrepancy with our analytical results and with the recent numerical results:
the RSPT predictions for $h_{c1}$ and $h_{c2}$ are concave up as functions
of $1/S$ where they should be concave down, and conversely, while the
non-normalized plateau width trends in the wrong direction. Thus only
in the extreme classical limit do the considerations of RSPT appear to
be valid.

In summary, we have investigated the 1/3 magnetization plateau of the
kagome antiferromagnet by using a straightforward modified spin-wave
theory, which contains only one self-consistent parameter. We have shown
that the quantum corrections contained in this magnon-density parameter
open finite energy gaps, which stabilize the $1/3$-plateau phase over a
broad range of magnetic fields. The qualitative and quantitative behavior
of the critical fields and plateau widths is in excellent agreement with
the recent numerical results for the same quantities obtained by exact
diagonalization \cite{ED_data} and by tensor-network methods \cite{iPEPS}.
These results indicate that the self-consistent spin-wave theory provides
an accurate description of the properties of the magnetization plateau in
the kagome antiferromagnet. We suggest that the same type of theory should
also be applied to describe the properties of magnetization plateaus in a
number of other frustrated systems, including the extended square and
honeycomb geometries as well as the triangular \cite{Triangular},
checkerboard, Shastry-Sutherland, and Husimi antiferromagnets \cite{Liao+16}.

This work was supported by the National Natural Science Foundation of China
(Grant Nos.~10934008, 10874215, and 11174365) and by the National Basic
Research Program of China (Grant Nos.~2012CB921704 and 2011CB309703).

\end{document}